\documentclass[%
reprint,
superscriptaddress,
 amsmath,amssymb,
 aps,
prb,
]{revtex4-2}

\usepackage{graphicx}
\usepackage{dcolumn}
\usepackage{bm}
\usepackage{hyperref}
\usepackage{sidecap}
\usepackage{float}
\usepackage{verbatim}
\usepackage{color}

\begin{document}
\title{Raman scattering of plane--wave and twisted light off chiral molecular liquids}

\author{Florian B\"uscher}
\affiliation{Institute for Condensed Matter Physics, TU Braunschweig, D-38106 Braunschweig, Germany}

\author{Silvia M\"ullner}
\affiliation{Institute for Condensed Matter Physics, TU Braunschweig, D-38106 Braunschweig, Germany}

\author{Dirk Wulferding}
\affiliation{Institute for Condensed Matter Physics, TU Braunschweig, D-38106 Braunschweig, Germany}
\affiliation{Center for Correlated Electron Systems, Institute for Basic Science,
	Seoul 08826, Republic of Korea}

\author{Yu. G. Pashkevich}
\affiliation{O.O. Galkin Donetsk Inst. for Physics and Engineering, NASU, Kyiv - Kharkiv 03028, Ukraine}
\affiliation{Institute for Condensed Matter Physics, TU Braunschweig, D-38106 Braunschweig, Germany}

\author{V. Gnezdilov}
\affiliation{B. Verkin Inst. for Low Temperature Physics and Engineering, NASU, 61103 Kharkiv, Ukraine}
\affiliation{Institute for Condensed Matter Physics, TU Braunschweig, D-38106 Braunschweig, Germany}

\author{A. A. Peshkov}
\affiliation{Institut f\"ur Mathematische Physik, TU Braunschweig, D-38106 Braunschweig, Germany}
\affiliation{Physikalisch-Technische Bundesanstalt, D-38116 Braunschweig, Germany}

\author{A. Surzhykov}
\affiliation{Institut f\"ur Mathematische Physik, TU Braunschweig, D-38106 Braunschweig, Germany}
\affiliation{Physikalisch-Technische Bundesanstalt, D-38116 Braunschweig, Germany}

\author{Peter Lemmens}
\affiliation{Institute for Condensed Matter Physics, TU Braunschweig, D-38106 Braunschweig, Germany}
\affiliation{Laboratory of Emerging Nanometrology LENA, D-38106 Braunschweig, Germany}

\date{\today}

\begin{abstract}
We present an experimental study of the quasi--elastic Raman scattering (QES) of plane--wave-- and twisted light by liquid crystals. Depending on their temperature, these crystals exhibit isotropic, nematic and chiral nematic phases. The question is addressed of how the phase of a crystal and the state of incident light can affect the quasi--elastic energy spectra of the scattered radiation, whose shape is usually described by the combination of Lorentzian and Gaussian components. Special attention is paid to the \textit{chiral phase}, for which the Raman QES spectrum is dominated by a Lorentzian with reduced linewidth, pointing to diminished disorder and configurational entropy. Moreover, this chiral phase is also known for a regime of iridescence (selective backscattering) which arises when the wavelength of incident light becomes comparable with the chiral pitch length. Detailed measurements, performed in this \textit{resonant} regime and by employing twisted light, carrying various projections of the orbital angular momentum (OAM), have indicated a low-energy scattering surplus depending on OAM. We argue that this observation might indicate a transfer of angular momentum between light and liquid crystal.



\end{abstract}

\maketitle

\section{Introduction}

Being a direct manifestation of mirror asymmetry, chirality plays an important role in chemistry, biology, and physics \cite{Anderson-72}. In particular, the concept of chirality has proven to be useful in explaining certain phenomena in molecules\cite{Chemical-Topology}, plasmonics\cite{Chiral-plasmonics}, quantum optics\cite{Chiral-quantum-optics}, Weyl- or Dirac semimetals\cite{Dirac-materials}, and enantiomers for life science\cite{Chirality-Sensing}. Chirality is intimately coupled to topology as they both refer to non-local properties that can be characterized by a conserved quantity, e.g. a Berry phase of Bloch electron states, depending themselves on symmetry.


During recent years, a particular interest in investigating topological phenomena in optics has arisen. One of the important testbeds for these studies is the twisted (or vortex) light, that possesses a helical phase front and carries a non--zero projection of the orbital angular momentum (OAM) onto its propagation direction \cite{Review-11,roadmap}. Among several ``twisted'' solutions of Maxwell equations, the Laguerre--Gaussian ones are in the special focus of experimental and theoretical studies. These modes have been intensively discussed and used, e.g. for enhanced and cryptography relevant information transfer \cite{Krenn-14}, to probe dark matter by extragalactic birefringence \cite{Minami-20, Castelvecchi-20, Rini-20}, as well as in life-science related superresolution microscopy and optical tweezers \cite{Review-11}. Twisted light can also transfer quantized angular momenta to Bose--Einstein condensates \cite{Ryu-07, Wright-13, Yakimenko-15}, quantum dots and single trapped ions \cite{Schmiegelow-16}.

The influence of OAM on the response of chiral objects to twisted light has meanwhile become a very controversial issue. According to theoretical studies \cite{Babiker-02, Andrews-04}, internal transitions in chiral molecules do not show any explicit dependence on the OAM of incident light. Indeed, neither an OAM specific circular dichroism \cite{Araoka-05, loeffler2011c} nor a Hermite-Gauss mode rotation have been observed in cholesteric liquid crystals \cite{loeffler2011b}. However, there are also indications in the literature that the optical OAM can engage with the chirality of a molecular system through higher-order processes \cite{Forbes-19b,Forbes-19c} or spin-orbit coupling \cite{Forbes-19}. Moreover, in standing wave optical configurations leading to superchiral light an enhanced signal of chiral molecules has been confirmed \cite{Tang-11,Rosales-12}.

This case might be different for Raman scattering as the latter involves different terms of light matter interactions. A detailed theoretical investigation of the Rayleigh and Raman scattering of twisted light by atoms and molecules has been recently presented in Refs.~\cite{Forbes-19c,Forbes-20}. This study is based on a Kramers--Heisenberg dispersion formula and accounts for the higher, non--electric--dipole terms, in the electron--photon interaction operator. It was shown, in particular, that the electric quadrupole (E2) coupling is very sensitive to the phase structure of incident twisted light, which opens up a possibility for the \textit{chiral} light--matter interactions. One can expect, therefore, that the use of twisted light beams, carrying different OAM's, could outperform the sensitivity of traditional circular dichroism. This would open up a completely new scheme of characterization and analysis of chiral matter using twisted light.

In order to further contribute to the discussion on the chiral light--matter interactions, we present here an experimental study of the quasi--elastic Raman scattering (QES) by liquid crystals. The advantage of such crystals is that they allow to investigate a wide range of physical phases, including the \textit{chiral} phase \cite{solladi-84,chirality-in-liquid-crystals}. The important characteristic of this phase is the pitch length that can be tuned by temperature, external electric and magnetic fields as well as by anchoring interfaces\cite{chirality-in-liquid-crystals}. There exist a huge optical rotation as well as other properties that make the chiral phase of liquid crystals an ideal building block for optoelectronics, including photonic band gap materials and lasers \cite{Xiang-16, Li-16, Yang-06, Balamurugan-16,Ozaki-04}.

Quasi--elastic Raman scattering is a well established technique to characterize liquid crystals. This is due to the large optical anisotropy and pronounced orientational fluctuations of the constituting molecules\cite{Chandrasekhar-76}. In general, light scattering and QES follow the fluctuation-dissipation theorem and their intensity is derived from an electronic susceptibility. There exist experimental studies of different phases \cite{Spector-95} as well as of confinement effects as function of film thickness and encapsulation \cite{Shalaginov-96,Stallinga-96,Bellini-95}. In ionic liquid crystals with larger intermolecular interactions, QES is found to reflect the short-time intermolecular dynamics. There exists a linear dependence of the QES intensity on the configurational entropy \cite{Urahata-06, Ribeiro-07, Ribeiro-11, Lima-19}. In the following, we will investigate the interaction of liquid crystals, being in chiral phase, with the Laguerre--Gaussian light modes. It will be shown that the energy spectra of scattered Raman radiation may be sensitive to the OAM projection of incident photons, if their wavelength become comparable with the pitch length of the crystal. We argue that this OAM--dependent effect might be an indication of the chiral light--matter coupling, discussed above.

\section{Experimental details}

In our Raman scattering (RS) measurements the incident light was generated with a Nd:YAG solid state laser (Compass 315M-100, Coherent) operating at a wavelength of $\lambda$=532.5~nm, a laser power of P=6.5~mW, as well as a $\lambda/4$-plate. The original right- (left-)hand circularly polarized plane-wave laser field was converted to Laguerre--Gaussian twisted radiation with OAM $l$=$+1$ ($-1$) using a voltage-controlled, variable spiral plate (Q-plate, Arcoptix) in transmission geometry. The helical wavefront structure of twisted light has been characterized using a Mach-Zehnder interferometer. The input twisted beam had an angle of incidence of about 30~$^\circ$ with respect to the normal of the sample surface. The scattered light was measured with a Dilor XY Raman Spectrometer in the frequency range from $\pm$20 to $\pm$720~cm$^{-1}$ (Stokes and anti-Stokes regime) using quasi--backscattering geometry and integrating with a lens of f=1.6 aperture.

As a test we performed scattering experiments on the surface of commercial silicon, representing a non-topological surface. Measurements were made as a function of focusing, number of optical elements, and integrated power. In no case did twisted light alter or modify the selection rules and scattering intensities of this surface. As the respective incident intensities differ for different polarization states, we have corrected the measured data to this intensity.

Of interest here is the light scattering by two different liquid crystal model systems which provide nematic, chiral, isotropic liquid, and solid phases as a function of temperature and therefore allow probing all aspects of topological light-matter interaction. Quartz glass cuvettes (size 45x12.5x12.5~mm$^3$ and a total sample thickness of 10~mm) with the samples were mounted on a Peltier heating/cooling stage (-20 to +100$^\circ$C) ensuring reasonable thermal contact. We used sequential heating and cooling cycles through the phase diagram to monitor the intensities of QES.

First, we analyzed the nematic 4’-Pentyl-biphenylcarbonitrile (Sample A, composition $CH_3(CH_2)_4C_6H_4C_6H_4CN$) using initial temperature scans in the range from 0 - 43$^\circ$C by first heating and then cooling with an equilibration time of one minute after the stabilization of the temperature. We observe that Sample A has two liquid phases and shows transitions from the crystalline to the nematic phase and from the nematic to the isotropic liquid phase at 18$^\circ$C and at 35$^\circ$C, respectively \cite{Aldrich-A}. This is based on the comparably simple elongated shape of the constituting molecules with two benzene rings that are corner bound. None of its phases show iridescent light scattering (selective reflection).

Next, we studied the cholesteryl nonanoate liquid crystal (Sample B, composition $C_{36}H_{62}O_2$), an ester of cholesterol and nonanoic acid. Initial heating scans of Sample B were performed by first heating the sample to 100$^\circ$C to melt the powder samples and then starting with a cooling scan in the temperature range between 20 - 100$^\circ$C. We noticed a consecutive degradation of Sample B following a larger numbers of cooling/heating cycles. This degradation shows up as an attenuation of the characteristic effects at the phase transitions and a shift of the phase transitions to higher temperatures. The properties of Sample B and our observations are summarized in Table I. As seen from this Table, Sample B shows four phases that appear with increasing temperatures as crystalline, smectic, chiral/cholesteric, and isotropic liquid. The transition temperatures are 77.5$^\circ$C, 79$^\circ$C, and 90$^\circ$C, respectively \cite{Aldrich-B}. The sequence of these phases is a direct consequence of the complex 3D bending of the four edge-sharing benzene rings that constitute the molecules, like a twisted banana. We observe, moreover, a pronounced iridescent light scattering with a bright orange color at T$_0$=84$^\circ$C. This selective reflection corresponds to a resonance of the temperature dependent, chiral pitch length with the wavelength of light \cite{Adamski-74}. It allows an independent assignment of the onset of its chiral phase. We will see below that this resonance, attributed to the chiral pitch length, might represent the condition under which the frequency spectrum of scattered light is affected by the ``twistedness'' of the incident radiation.

\section{Results}

To determine the lineshape and width of the QES we have performed combined Stokes and anti-Stokes measurements at several characteristic temperatures associated with the different phases as described above. More detailed intensity data were obtained using only Stokes RS. In the inset of Fig.~1 we present QES spectra for scattering of linear polarized, plane-wave radiation off nematic Pentyl-biphenylcarbonitrile (Sample A) with linear polarized, plane--wave radiation. There exist a Gaussian QES maximum with a linewidth of FWHM=100~cm$^{-1}$ and two shoulders on the Stokes and anti--Stokes sides, at $E$=$\pm$89~cm$^{-1}$, and a linewidth of FWHM=50~cm$^{-1}$. The the isotropic, liquid phase of Sample A shows an additional fluorescence signal as a background that is quenched in all anisotropic phases of the liquid crystals. The characteristic molecular vibrational modes are at higher energies and will not be discussed here \cite{Aldrich-A}.

\begin{figure*}
	\label{figure1}
	\centering
	\includegraphics[width=16cm]{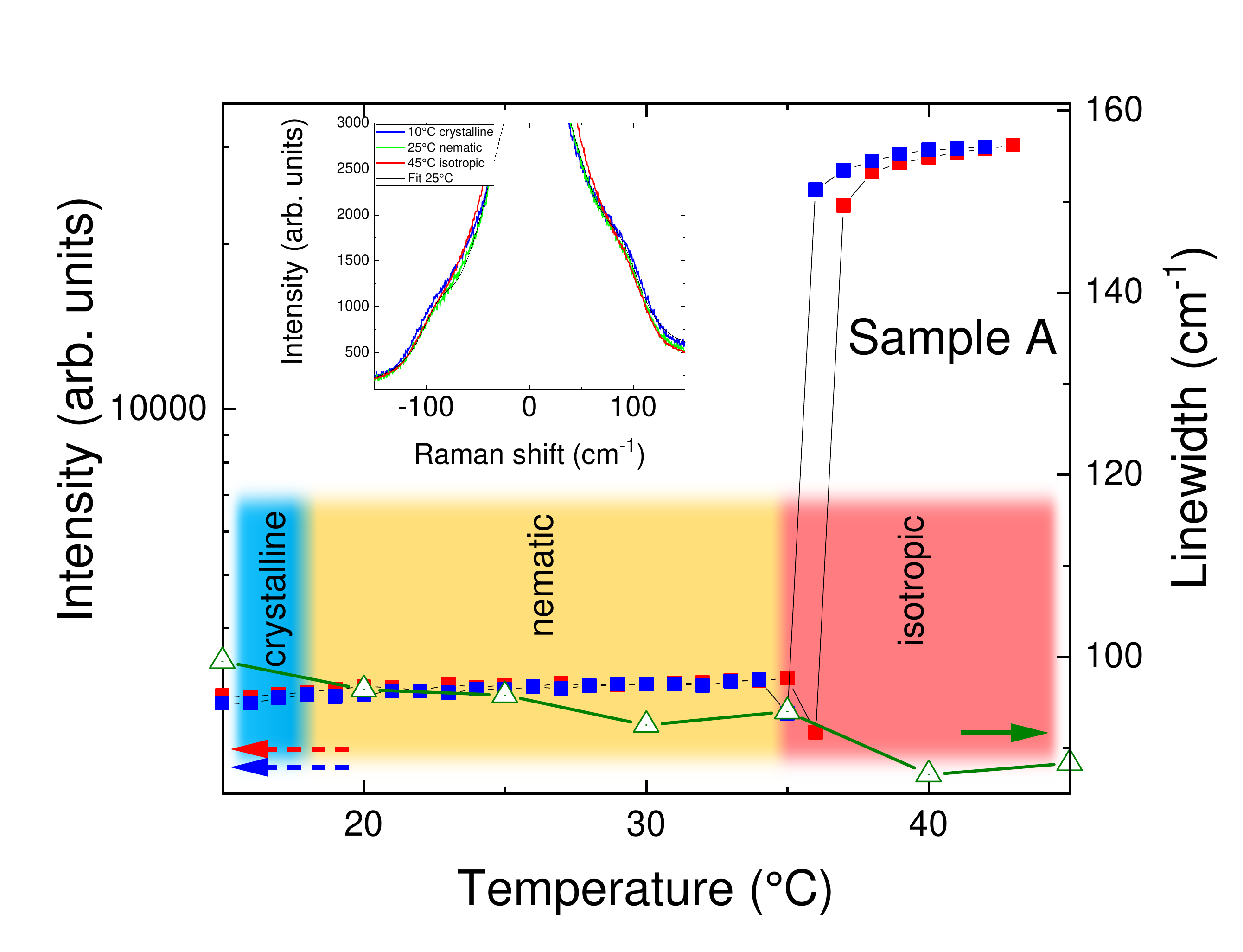}
	\caption{(Color online) Integrated intensity and linewidth (FWHM) of the QES spectrum, measured for the nematic Pentyl-biphenylcarbonitrile (Sample A). The integrated intensity is shown on a logarithmic scale for heating (red squares) and cooling (blue squares), respectively. The linewidth data (FWHM, green triangles) are determined from fitting combined Stokes-anti-Stokes spectra. The inset shows spectra of Raman scattered light, measured at certain temperatures, that represent all phases of the liquid crystal.}
\end{figure*}

As seen from Fig.~1, there is a pronounced effect of temperature on the QES intensity, integrated over the frequency of scattered light. Namely, if the sample is cooled down, this integrated intensity is first gradually reduced for $35^\circ \lesssim T < 50^\circ$ and then suddenly drops to a much smaller value. This drop of intensity is attributed to the quenching of fluorescence.

By studying the variation of the integrated intensity for cooling down and warming up the crystal sample, we observed a hysteresis of approximately 1$^\circ$C at the temperature $T \approx 35^\circ$, that marks a transition between isotropic and the nematic phases. The finite energy Lorentzian at $\pm$89~cm$^{-1}$ shows a similar drop of intensity while its linewidth broadens from 50 to 55~cm$^{-1}$ at 30$^\circ$C, not shown here.

In contrast to the integrated intensity, the linewidth of QES, displayed by the green line in Fig.~1, is only moderately affected by temperature. Indeed, the linewidth, measured for the isotropic liquid phase at high temperatures $T \gtrsim 35^\circ$, is only about 10\% smaller compared to the linewidth in the nematic and 20\% smaller compared to the crystalline phase. The corresponding data points are derived from combined Stokes-anti-Stokes spectra and therefore their density is not so high.

In Fig.~2 we show the corresponding intensity and linewidth data for the cholesteryl nonanoate liquid crystal (Sample B). In contrast to the Sample A, the QES energy spectrum exhibits a Lorentzian line-shape and no evidence for shoulders exists at finite energies, see inset of Fig.~2. The linewidth (green line) and integrated intensity (blue and red lines) of this mode show a pronounced temperature evolution including a hysteresis. In particular, the linewidth drops from approximately 80 to 10~cm$^{-1}$ at the transition from isotropic to the chiral phase which opposes a behaviour of the linewidth in Sample A. The QES intensity drops by nearly two orders of magnitude in this temperature range and forms a weak minimum at $T \approx$ 95~K. We attribute this minimum to the onset of the chiral phase. Further minima and maxima of the integrated intensity can be observed at 80(85)$^\circ$ and 76(78)$^\circ$, respectively. These temperatures correspond to a cooling (heating) of the sample and are identical to the boundaries to the smectic and crystalline phases known from literature \cite{Aldrich-B}.

In general, the cooling data, displayed by blue curve, show more defined changes at the characteristic temperatures. We attribute this to a smaller configurational disorder in Sample B with heating the sample after the initial melting and cooling down. The onsets of the isotropic and the smectic phases show a pronounced hysteresis of cooling/heating data of about 4$^\circ$C. We attribute these to a weaker pinning of domain walls due to a reduced configurational disorder. Moreover, as mentioned above, also the Lorentzian linewidth of QES in Sample B of the chiral and nematic phases is much smaller than that of the Gaussian linewidth in Sample A.

\begin{figure*}
	\label{figure2}
	\centering
	\includegraphics[width=16cm]{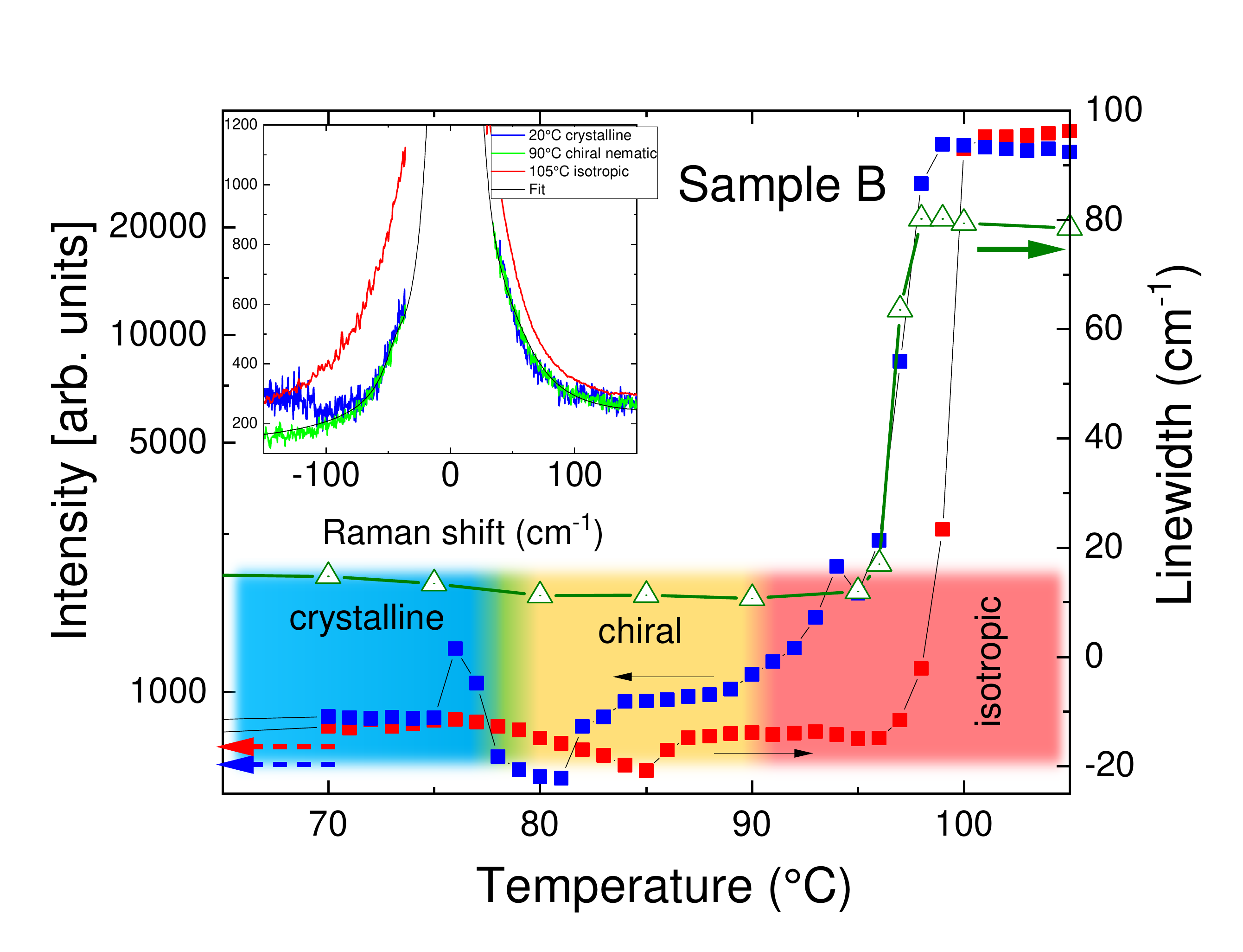}
	\caption{(Color online) Intensity and linewidth (FWHM) of the QES from cholesteryl nonanoate liquid crystal  (Sample B). The intensity is shown on a logarithmic scale for heating (red squares) and cooling (blue squares). The linewidth data (triangles) are determined from fitting combined Stokes and anti-Stokes spectra. The inset shows representative spectra from all phases.}
\end{figure*}

Table I summarizes the observations at the characteristic temperatures of the chiral liquid crystal (Sample B). The biggest difference compared to the nematic (Sample A) liquid crystal is given by the Lorentzian lineshape of its QES. We relate this to the difference between disorder dominated versus fluctuation dominated processes. Moreover, as mentioned already above, the temperature behaviour of the linewidth of the spectrum differs qualitatively for Samples A and B. As seen from the table and Fig.~2, a significant reduction of the Sample B linewidth is observed for cooling the target crystal to $T < 95^\circ$, i.e. into the anisotropic liquid phases.

\begin{table*}
	\caption{\label{tab:table2} Properties and characteristic temperatures of the cholesteryl nonanoate liquid crystal (Sample B). The temperatures, that mark transitions between different phases, are taken from literature \cite{Aldrich-B}. The described effects in RS data correspond to cooling (heating) the sample across the transitions. }

		\begin{tabular}{ c|c|c|c|c|c|c|c }\hline
 Cholesteryl nonanoate (Sample B) & cryst. &         & smectic &      	& chiral  &      & iso. liquid \\ \hline
T$_c$ \cite{Aldrich-B} [$^\circ$C]	&	& 77.5  &	       &  79	&	        &  90  &		-     \\\hline
Lorentzian linewidth [cm$^{-1}$]&20  	 &	     &   18    &		&   16	    &	     &      80     \\
Linewidth anomalies		        &	     & slope &         &  		&	        & step &             \\ \hline
Intensity [arb. units]      	&  850	 &	     &1300(830)&        &1000(750)  &      & 33000(36000)\\
Intensity anomalies			    &		 &76(78) &		   & 80(85)	&  		    &96(99)&		-       \\ \hline
Iridescence, $T_0$ [K]			&		 &		 & 		   &  84 	&           &      &            \\ \hline
	\end{tabular}
\end{table*}


The pronounced anomalies such as peaks, broadened maxima, and minima in the temperature dependence of the QES intensity indicate that the Raman scattering provides a sensitive probe of the intrinsic fluctuation spectra of a liquid crystal. The general qualitative agreement between the characteristic temperatures, observed in this study, and data from literature \cite{Aldrich-B} supports their assignment to the phase boundaries of the molecular liquids. We attribute also the hysteresis effects to a microscopic origin, i.e. first order phase transitions separating the phases as well as the strong dependence of the chiral phase with rather extended pitch lengths, in particular, on thermal fluctuations and stress induced disorder.

Despite a rather good general agreement between the characteristic temperatures, measured in this study and reported in the literature, one can observe some quantitative discrepancies. For example the phase boundary between the chiral and the isotropic phase seems to be shifted to higher temperatures. One can mention several possible reasons for these discrepancies. First of all, there exists a temperature mismatch of a few degrees between the thermometer at the heating/cooling stage and the sample in the laser spot. More importantly, the surface of the cuvette affects the liquid molecule orientation in its proximity where the RS data are gained from. Such an interface acts as a director of orientation and reduces thermally induced orientational fluctuations. This shifts the observed phase boundaries and characteristic anomalies to higher temperatures \cite{Napoli-12}. Therefore, we use the observation of iridescent scattering (orange colour) for Sample B at the nominal stage temperature of $T_0$=84$^\circ$C as a kind of independent fixpoint of the chiral phase.

Until now we have analyzed the Raman spectra by studying their intensities and linewidths. Indeed, such an analysis strongly relies on the assumption about a particular lineshape. The detection of contributions with unknown characteristics and small intensity is therefore limited. Therefore, we will continue our study using a different approach in which the experimental findings will be \textit{scaled} by dividing spectral intensities of scattered radiation, obtained for different states of incident light or different phases of a target crystal. This approach will allow to better understand the sensitivity of the Raman scattering on polarization and even OAM state of (initial--state) light. The other, experimental, advantage of the ``scaling'' approach is that it reduces the effect of the different numbers of optical elements involved in the scaled data. This approach is similar to Raman optical activity (ROA), where rather small changes of scattering intensities are amplified with respect to unspecific background shifts.



In Fig.~3 we present normalized spectra given by the ratio of the spectral intensities of scattered light observed for incident linearly polarized plane waves to that observed for incident circularly polarized, twisted light. Results are displayed for different phases of the two molecular liquids. For sample A, we see a flat signal of the order of unity, denoted by a colored marker on the right axis. The observed noise is especially small at low energies where the underlying QES data has higher intensity. A small deviation from this marker in the high frequency limit is attributed to the variation of incident laser intensity. However, a quite different pattern is revealed for sample B. Namely, there is an increasing low-energy surplus (green line), which corresponds to the chiral nematic phase.

\begin{figure*}
	\label{figure3}
	\centering
	\includegraphics[width=12cm]{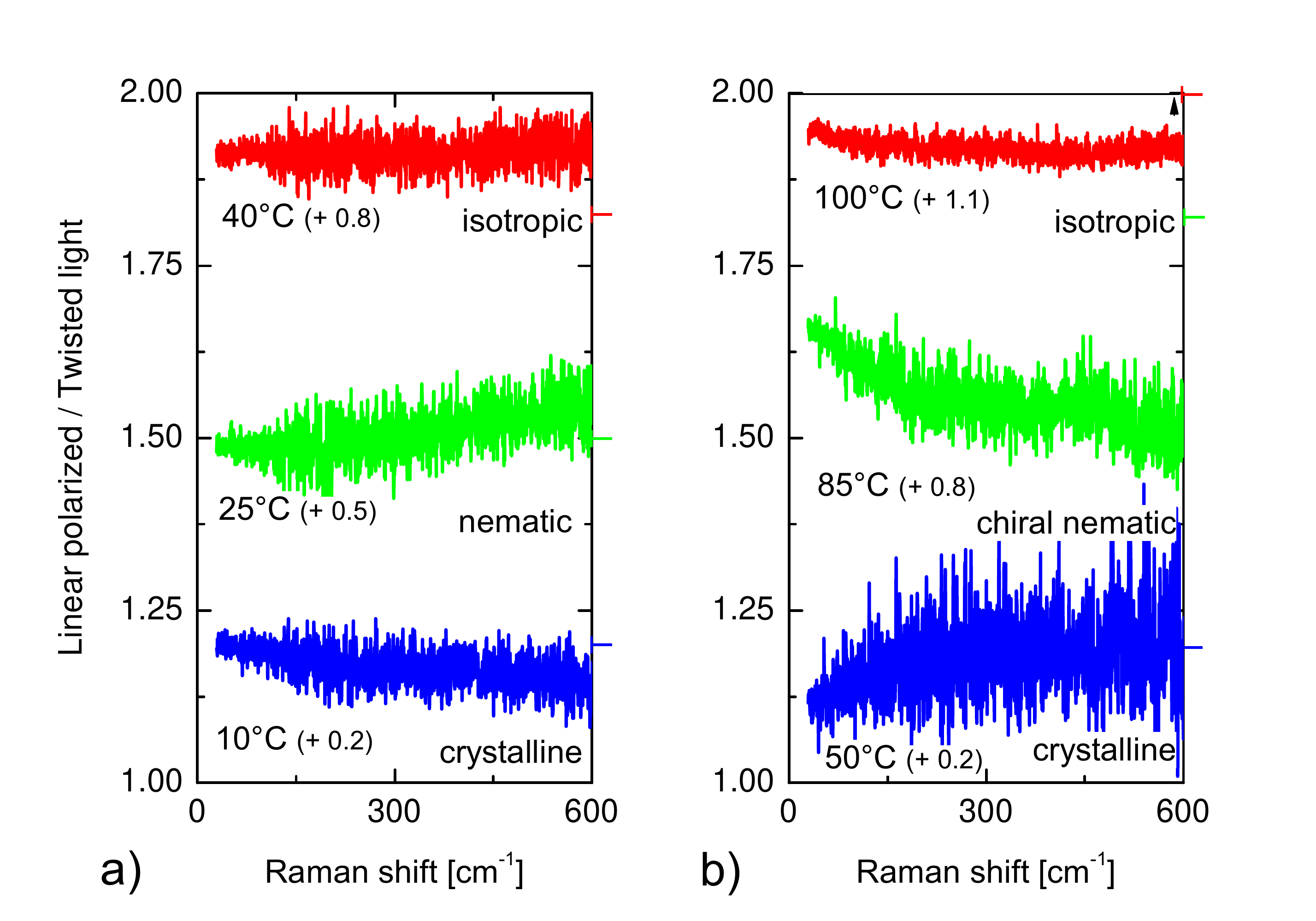}
	\caption{(Color online) Normalized RS spectra of Sample A in the left panel and Sample B in the right panel for three different phases of the target liquid crystal, tuned-in by temperature.
	To perform a normalization procedure, we calculated the ratio of the spectral intensities of scattered light, obtained for incident linearly polarized plane--wave and circularly polarized, twisted radiation. For the latter, we employed the Laguerre--Gaussian modes with the OAM $l$=$+1$. The resulting graphs have been shifted for convenience, see number in brackets. Colored markers on the right hand side correspond to respective unity of normalization. }
\end{figure*}

\begin{figure*}
	\label{figure4}
	\centering
	\includegraphics[width=12cm]{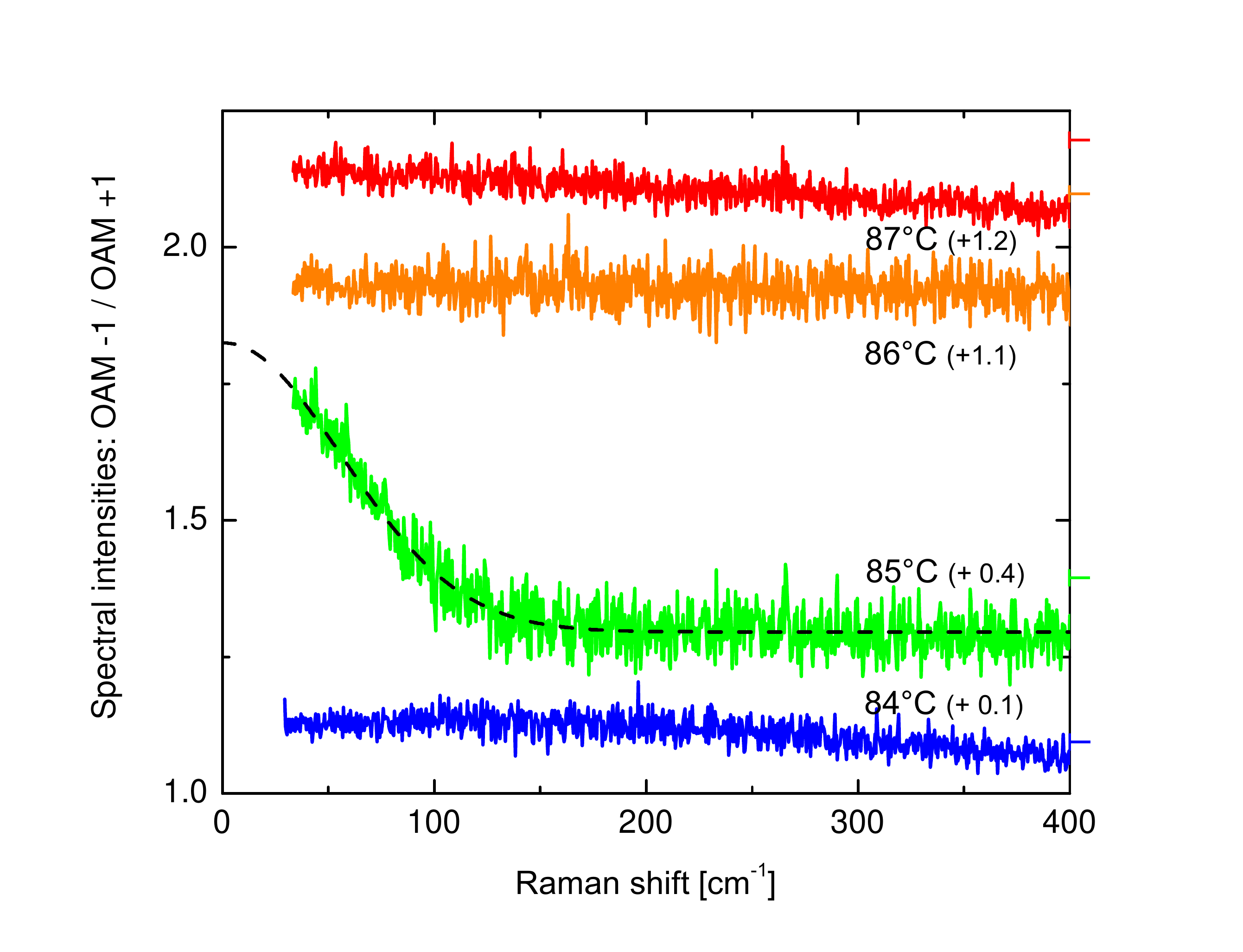}
	\caption{(Color online) The ratio of the spectral intensities of scattered light, measured for incident Laguerre--Gaussian twisted radiation with OAM $l$=$\pm1$. The dashed line corresponds to a fit by a Gaussian line with the center x$_0$=0~cm$^{-1}$ and the linewidth $\Gamma$=130~cm$^{-1}$. The resulting graphs have been shifted for convenience, see number in brackets. Colored markers on the right hand side correspond to respective unity of normalization. }
\end{figure*}

In order to understand better the OAM dependence of the Raman scattering, in Fig.~4 we show show the spectral intensities of outgoing radiation obtained for incident twisted light with OAM $l$=$-1$ divided by that for $l$=$+1$. Such normalized RS spectra are displayed only for Sample B within the chiral nematic phase. Here, the small temperature steps were chosen around iridescence at orange, T$_0$=84$^\circ$C. We find that increasing the temperature to T$_0$+1$^\circ$ and T$_0$+2$^\circ$C shifts iridescence to green and to blue. We also see that the signal at T=T$_0$, as well as at T$_0$+2$^\circ$C and higher temperatures, is very close to unity, thus revealing no significant effect of OAM. Small deviations of the data from unity marked by a colored tick on the right hand axis are attributed to a variation of the incident laser intensity with the number of optical elements. On the other hand, a pronounced difference in scattering intensity between scenarios with OAM $l$=$+1$ and $l$=$-1$ is observed for T=T$_0$+1$^\circ$ and can be modeled by Gaussian quasi--elastic line, see fit in Fig.~4. This low energy scattering surplus, obtained for the resonant conditions given by iridescence, might indicate the influence of optical OAM on the Raman scattering within the chiral nematic phase.





\section{Discussion}

A comparison of the QES intensity of nematic and chiral phases of Sample A and B in Figs.~1 and 2, respectively, shows, that well defined intensity anomalies can only be identified for Sample B with QES of Lorentzian lineshape and temperatures below $\approx$~95~K. Here, the linewidth is rather small and we interpret the QES as being dominated by intrinsic Lorentzian fluctuations. Such data can be used to characterize all topological phases and study, e.g. ordering phenomena and hysteresis effects, see for comparison QES of a Kagome spin liquid \cite{kagome}. In contrast, all phases of Sample A are dominated by defects leading to a Gaussian lineshape of larger linewidth, usually related to diffusion processes.

In agreement to the available literature, moreover, we do not find the dependence of the Raman scattering by chiral molecules on the OAM of incident twisted light. This statement is based on the negligible, low energy upturn of scaled intensity for the chiral phase of Sample B in Fig.~3b. The observed enhancement at $\Delta\omega$=50~cm$^{-1}$ is smaller than I$_{lin}$/I$_{l=+1}$=(10$\pm$3)$\%$ with considerable background noise.


A much larger surplus of the (near--elastic) scattering is observed in Fig.~4, where we display the ratio of the spectra, obtained for incident Laguerre--Gaussian modes with OAM projections $l$=$\pm 1$. For the Raman shifts below 50~cm$^{-1}$, for example, we find the spectral intensity ratio I$_{l-1}$/I$_{l+1}$=(35$\pm$3)$\%$, which may indicate a dependence of the Raman scattering on the OAM. The very small temperature interval of this effect points to a resonance-like origin. Indeed, for the temperature of the sample of $T \approx 85^\circ$, the pitch length of the chiral phase becomes comparable with the wavelength of incident light.

In order to provide (at least qualitative) explanation of the observed OAM--dependent effect, we remind that according to the fluctuation--dissipation theorem, the intensity of the Raman scattered light can be obtained as:
\begin{equation}\label{RamanInt}
I_R\sim \omega_I\omega_S^{3}\int
\ll\chi(\vec{k}_I\vec{k}_S;t)\chi^{\dagger}(\vec{k}_I\vec{k}_S;0)\gg
e^{i\omega t}dt,
\end{equation}
where $\chi$ are electronic susceptibilities that depend on polarization-- and wave--vectors of incident and scattered light. The evaluation of these susceptibilities can be traced back to the matrix elements that describe molecular (electronic, vibrational and rotational) transitions induced by light. As usual in atomic and molecular structure calculations, these transition matrix elements can be expanded in terms of (electric and magnetic) multipole terms. It is well known that account for the electric dipole (E1), electric quadrupole (E2) and magnetic dipole (M1) is important for the proper description of the optical activity of target molecules, i.e. their different response to the left- and right circularly polarized light. Moreover, it was recently discussed that E2 coupling might be also crucial for the description of the influence of the orbital angular momentum of light on the scattering process \cite{Forbes-19c,Forbes-20,Forbes-21}. Since the role of the higher-order contributions increases when the wavelength of incident light becomes comparable to the pitch length of the liquid crystal in the chiral phase, this can be an explanation for the observed effect.

The experimental data of lineshape and linewidth (FWHM) of the observed spectra can provide further information about the OAM dependence of the Raman scattering by chiral liquid crystals. To perform this analysis we note that the scattering surplus fits well to a Gaussian distribution, centered at E~$\approx$~0 and having the width $\Gamma$$\approx$130~cm$^{-1}$. This implies that the underlying process is diffusive and has no lower boundary or energy gap. The characteristic upper energy limit of 170~cm$^{-1}$$\approx$250~K is comparable to the thermal energy of the phase with T$_{o}$=360~K. The observed $\Gamma$ is roughly one order of magnitude larger than the original Lorentzian QES linewidth (16~cm$^{-1}$) of the chiral phase probed by linear polarized plane wave. This suggests that the OAM transfer observed with twisted light is definitely different from the QES of Lorentzian lineshape that is observed as a dominant contribution to linear polarized light. 	

Indeed, an OAM transfer is not expected to lead to a considerable energy transfer to the molecular system since the related intermolecular energy scales are rather small. This can be estimated from the related difference $\Delta$T$_{c,i}$/T$_c \approx$ 0.4\% comparing the smectic with the chiral phases. Therefore, the scattering process is expected to be quasi-elastic. The dynamics of the molecules themselves is based on diffusion and drift. This leads to an Gaussian envelope as a typical solution of stochastic processes described within Fokker-Planck and Langevin equations \cite{Haken-83}. Interestingly, this also negates the existence of a collective mode of OAM transfer.

\section{Summary}

The observed low energy Raman scattering response of liquid crystals is dominated by QES whose intensity can be used to characterize their topological phases. While liquid crystals with nematic phases show QES with large linewidth attributed to a domination of disorder, the chiral cholesteryl nonanoate shows QES of smaller linewidth and Lorentzian lineshape.

Despite a controversial discussion about the sensitivity of the photon scattering by chiral matter on the OAM of incident twisted light, several recent studies have suggested that certain conditions for such an OAM coupling can be anticipated. These are electronic resonances, confinement, and phase coherent superpositions that are advantageous for higher--order terms of the vortex--light--matter coupling. Our experiments suggest that the observation of iridescence in the chiral phase of a liquid crystal may help to investigate these regimes, favorable for the OAM transfer between light and matter. Based on our observations we argue that the resulting scattering dynamics is Gaussian and diffusive, without evidence for collective mode and has a rather large linewidth comparable to the thermal energy within this phase. Therefore, the Raman scattering experiments could open up an avenue to a better understanding of light--matter interaction involving twisted light.

%

\begin{acknowledgments}
We acknowledge important discussions with W. L\"offler (Univ. Leiden) and G. Napoli (Univ. del Salento, Lecce). This research was funded by DFG EXC-2123 QuantumFrontiers – 390837967, DFG Le967/16-1, DFG-RTG 1952/1, and the Quantum- and Nano-Metrology (QUANOMET) initiative of Lower Saxony within project NL-4. D. W. acknowledges support from the Institute for Basic Science (Grant No. IBS-R009-Y3).
\end{acknowledgments}


\end{document}